\documentclass[a4paper,11pt]{article}

\usepackage[latin1]{inputenc}
\usepackage[top=2.5cm, left=2.5cm, right=2cm, bottom=2.5cm]{geometry}
\usepackage{graphicx}
\usepackage[centertags]{amsmath}
\usepackage{amsfonts}
\usepackage{amssymb}
\usepackage{amsthm}
\usepackage{newlfont}
\usepackage{xcolor}
\usepackage{soul}
\usepackage{lscape}

\usepackage{xcolor} \definecolor{darkgreen}{rgb}{0,.5,0}
\usepackage[colorlinks,filecolor=blue,citecolor=darkgreen,unicode]{hyperref}

\begin{document}
\begin{center}

{\Large \bf{ Boundary Conditions for $\mathrm{AdS_2}$ Dilaton Gravity} }

\vspace{1cm}
Carlos Valc\'arcel\footnote{valcarcel.flores@gmail.com}

\vspace{.5cm}

$^{a}$\emph{Instituto de F\'isica, Universidade Federal da Bahia, C\^ampus Universit\'ario de Ondina, 40210-340, Salvador-BA, Brazil}.

\end{center}

\vspace{.25cm}

\begin{abstract}

We study a bi--parametric family of dilaton gravity models with constant and negative curvature. This family includes the Jackiw--Teitelboim gravity and the Liouville gravity model induced by a bosonic string. Furthermore, this family is conformally equivalent to the hiperbolic dilaton models. We propose boundary conditions in the Fefferman--Graham and in the Eddington--Finkelstein gauge. We check the consistency of the asymptotic conditions by computing the entropy of their black hole solution. 

\vspace{.5cm}


\end{abstract}

\section{Introduction}

The study of $\mathrm{AdS_2}$ holography began in the late nineties \cite{Strominger:1998yg,Cadoni:1998sg,Hotta:1998iq,NavarroSalas:1999up} exploring the asymptotic symmetries of the Jackiw--Teitelboim (JT) gravity \cite{Teitelboim:1983ux,Jackiw:1984je} and dilaton models that arise from dimensional reduction. In the last years, the interest for $\mathrm{AdS_2}$ holography was renewed due to the discovery of the holographic 
equivalence \cite{Maldacena:2016hyu,Jensen:2016pah} between JT gravity and the Sachdev--Ye--Kitaev (SYK) model \cite{Sachdev:1992fk,Kitaev} whose dynamics is described, in a certain limit, by a Schwarzian action \cite{Mertens:2018fds}. There has also been activity on the study of the duality between JT gravity, and other dilaton theories, with random matrix models \cite{Saad:2019lba,Witten:2020wvy,Turiaci:2020fjj}.

Another important aspect of JT gravity is that it describes a sector of the dynamics of near extremal black holes \cite{Maldacena:2016upp}. Then, several properties of JT gravity can be uplifted to higher dimensions. However, $\mathrm{AdS_2}$ holography is not limited to pure JT gravity. There have been several works exploring super-symmetric \cite{Astorino:2002bj,Forste:2017kwy,Cardenas:2018krd,Fan:2021wsb} and higher-spin \cite{Gonzalez:2018enk} extensions of JT 
as well as super--symmetric extensions of  Liouville gravity \cite{Mertens:2020hbs,Mertens:2020pfe}, two dimensional Einstein-Maxwell-Dilaton 
theory \cite{Cvetic:2016eiv} and the $ab$--family \cite{Katanaev:1997ni,Ecker:2021guy}. The study of flat--space holography 
\cite{Afshar:2019axx,Afshar:2020dth} is based on a twisted version of the Callan--Giddings--Harvey--Strominger (CGHS) gravity model  \cite{Callan:1992rs} 
but can also be analyzed from the JT gravity perspective \cite{Godet:2020xpk,Afshar:2021qvi}. There has also been recent activity exploring the 
non--relativistic limits of dilaton gravity \cite{Grumiller:2020elf,Gomis:2020wxp} and building new classes of the dilaton 
actions \cite{Grumiller:2021cwg,Ecker:2022vkr,Grumiller:2022poh}.

In this work we study Hotta's model \cite{Hotta:1998iq}, which consists in a bi--parametric family of two dimensional gravity theories 
with constant and negative curvature, called $\mathrm{AdS_2}$ dilaton gravity. This model is an interesting laboratory to explore lower dimensional holography: 
Since it has constant curvature $R=-2$ everywhere, not just asymptotically, it allow us to identify which properties of the $\mathrm{AdS}_2$ holography 
are related to JT gravity. Furthermore, the $\mathrm{AdS_2}$ dilaton gravity model is conformally related to the hyperbolic 
potentials \cite{Kyono:2017jtc,Kyono:2017pxs,Frolov:2017onj,Okumura:2018xbh} obtained from the Yang-Baxter deformation technique. 

Our objective is to propose consistent boundary conditions for this model. We achieve this by solving the equations of motion in the first--order formalism, analogous to \cite{Grumiller:2015vaa,Ecker:2021guy} for other dilaton models. We use Euclidean signature in order to study the thermodynamic properties of the theory and check the validity of our boundary conditions.
Our interest in working in this formalism is the following: 
We know that all dilaton theories can be written as a Poisson--sigma model (PSM) \cite{Schaller:1994es}, a topological theory. Then, PSM is an unifying 
framework to study general properties shared by all dilaton models. More recently, in \cite{Valcarcel:2022zqm} was showed that we can relate asymptotic 
symmetries of these models by performing a target space diffeomorphism, an special symmetry of the PSM. Therefore, by proposing boundary conditions in 
the first--order formalism for $\mathrm{AdS}_2$ dilaton models, we can map these conditions to other models related by target space diffeomorphisms. 

This paper is organized as follows. In section \ref{sec2} we build and study some properties of the $\mathrm{AdS}_2$ dilaton gravity. In section \ref{sec3}, 
we review the first--order formulation for dilaton gravity. In section \ref{sec4}, we solve the first-order equations of motion in the Fefferman--Graham gauge, 
we propose boundary conditions and compute the asymptotic symmetries. In section \ref{sec5}, we propose boundary conditions in the Eddington--Finkelstein 
gauge using complex fields. In section \ref{sec6}, we compute the entropy of the black hole solutions using Wald's formula. Finally, in section \ref{sec7} 
we comment our results and give future perspectives of this work.

\section{$\mathrm{AdS}_2$ dilaton gravity models}\label{sec2}

A large variety of two-dimensional dilaton gravity models \cite{Nojiri:2000ja,Grumiller:2002nm} are described by the following Euclidean action
\begin{equation}
I\left[X,g_{\mu\nu}\right]=-\frac{k}{4\pi}\int_{\mathcal{M}}\mathrm{d}^{2}x\;\sqrt{g}\left[XR-U\left(X\right)\left(\partial X\right)^{2}-2V\left(X\right)\right] \label{D01}
\end{equation}
where $\mathcal M$ is a two-dimensional manifold, $g_{\mu\nu}$ is the metric, $X$ is the dilaton field, $R$ the curvature, $U$, $V$ are dilaton potentials and $k$ the normalization factor. Different potentials $U$ and $V$ describe different gravity models. We can cite, for example, the $\mathrm{JT}$ gravity \cite{Teitelboim:1983ux,Jackiw:1984je} $\left(U=0,\,V=-\Lambda X\right)$ or the $\mathrm{CGHS}$ model \cite{Callan:1992rs} $\left(U=0,\,V=-\lambda/2\right)$. 

In \cite{Bergamin:2004pn} it was showed that all dilaton models \eqref{D01} allow non--constant dilaton solution
\begin{equation}
\mathrm{d}s^{2}=\frac{1}{e^{Q}\xi}\mathrm{d}r^{2}+e^{Q}\xi\mathrm{d}\theta^{2},\qquad\partial_{r}X=e^{-Q}     \label{d03}
\end{equation}
where $r$ is the radial coordinate and $\theta$ a coordinate. The functions $Q$, $w$ and $\xi$ are:
\begin{equation}
Q\left(X\right)=\int^{X}\mathrm{d}y\;U\left(y\right),\quad w\left(X\right)=\int^{X}\mathrm{d}y\;e^{Q\left(y\right)}V\left(y\right),\quad\xi\left(X\right)=2\mathcal{C}-2w\left(X\right)  . \label{d04}
\end{equation}
In \eqref{d04} we introduced the Casimir $\mathcal C$, a constant of motion. From the line element \eqref{d03} it is straigthforward to compute the curvature:
\begin{equation}\label{d05}
R = -e^{-Q}\left[\partial_{X}^{2}\xi+\xi\partial_{X}U+U\partial_{X}\xi\right].
\end{equation}
We are interested in building a family of dilaton models with constant curvature. Let us consider a constant potential $U=4\sigma^2$ and an unknown potential $V(X)$. Replacing these potentials in \eqref{d05} for $R=-2$, i.e., $\mathrm{AdS_{2}}$ space, we obtain a first-order partial differential equation: $\partial_{X}V+8\sigma^2 V=-1$ which can be easily solved. Therefore, the $\mathrm{AdS_{2}}$ gravity models are described by the following potentials \cite{Hotta:1998iq}:
\begin{equation}\label{d06}
U\left(X\right)=4\sigma^{2},\qquad V\left(X\right)=\frac{1}{8\sigma^{2}}\left(V_{0}e^{-8\sigma^{2}X}-1\right)    
\end{equation}
where $\sigma$, $V_{0}$ are two parameters. We can identify some interesting limits depending on the value of the parameters: For $V_0=1$ and $\sigma^2\rightarrow 0$, the potential $U$ goes to zero and $V\approx-X$. Then, the action reduces to $\mathrm{JT}$ gravity. For $V_0=0$, \eqref{d06} reduces to Liouville gravity induced by a bosonic string \cite{Polyakov:1981rd,Bergamin:2004pn}. For $\sigma=1/2$ and $V_0$ arbitrary, the model reduces to the Liouville gravity presented in \cite{Grumiller:2007wb} which arise from the spherical reduction of pure Einstein gravity in $2+\varepsilon$ dimensions as $\varepsilon$ goes to zero. 

The construction of $\mathrm{AdS_2}$ gravity models began with the choice of positive and constant value for the potential $U$. For $U = 0$, the construction lead us to the Almheiri--Polchinski model \cite{Almheiri:2014cka}. On the other hand, if we begin with negative values $U=-4\sigma^2$, we notice that the asymptotic region is located at $r\rightarrow\infty$. The latter case is ``non--standard" since the asymptotic region is located at $r\rightarrow -\infty$. For more general potentials $U$, the curvature will depend on the Casimir.

For the model \eqref{d06}, the linear dilaton equation $\partial_{r}X=e^{-4\sigma^{2}X}$ has solution
\begin{equation}
 X\left(r\right)=\frac{1}{4\sigma^{2}}\ln\left(4\sigma^{2}r\right)    
\end{equation}
for $\sigma \neq 0$. This means that the asymptotic region $r\rightarrow\infty$ corresponds to $X\rightarrow\infty$. The $g_{\theta\theta}$ component of the line-element \eqref{d03} is:
\begin{equation}
g_{\theta\theta} = e^{Q}\xi\left(r\right)=\frac{1}{16 \sigma^{4}}e^{8\sigma^{2}X}+2\mathcal{C}e^{4\sigma^{2}X}+\frac{V_{0}}{16\sigma^{4}}=r^{2}-\frac{c_{0}}{2\sigma^{2}}r+\frac{V_{0}}{16\sigma^{4}}    
\end{equation}
where we defined a re-scaled Casimir $c_{0}\equiv-16\sigma^{4}\mathcal{C}$. This can be interpreted as a black hole solution with outer and inner horizons $r_{\pm}$ located at: $e^{Q}\xi\left(r_{\pm}\right)=0$, where $r_{\pm} = \frac{c_{0}}{4\sigma^{2}}\pm\frac{1}{4\sigma^{2}}\sqrt{c_{0}^{2}-V_{0}}$. In order to have real and positive values for the outer horizon we need to impose that $c_{0}$ is positive and $c_{0}^{2}>V_{0}$. This means that the original Casimir $\mathcal{C}$ is negative. Black hole solutions of this kind have been studied in the context of gravity coupled with the trace of the two--dimensional energy momentum tensor \cite{Mann:1990gh}.

From regularity of the metric at the horizon we obtain the periodicity $\beta_\theta$ and the temperature $T$:
\begin{equation}
\beta_{\theta}=\frac{16\sigma^{2}\pi}{\sqrt{c_{0}^{2}-V_{0}}} ,\qquad T = \frac{\sqrt{c_{0}^{2}-V_{0}}}{16 \sigma^{2}\pi}.   
\end{equation}
We can now compute the entropy of the black hole, which is proportional to the dilaton evaluated at the horizon \cite{Gegenberg:1994pv}.
\begin{equation}
S_{\mathrm{Wald}}=kX\left(r_{+}\right)=\frac{k}{4 \sigma^{2}}\ln\left(4\sigma^{2}r_{+}\right)=\frac{k}{4\sigma^{2}}\ln\left(c_{0}+\sqrt{c_{0}^{2}-V_{0}}\right). \label{entropy}
\end{equation}
The logarithmic expression for the entropy is reminiscent of Liouville gravity. 

It is worth noticing that the $\mathrm{AdS}_2$ models \eqref{d06} are also conformally equivalent to hyperbolic dilaton potentials \cite{Kyono:2017jtc,Kyono:2017pxs,Mertens:2020hbs}. By performing a conformal transformation $\hat{g}_{\mu\nu}=e^{-4\sigma^{2}X}g_{\mu\nu}$, the new potential $\hat{U}$ vanish and $\hat{V}$ becomes
\begin{equation}\label{d07}
\hat{V}=-\frac{1}{8\sigma^{2}}\left[\left(1+V_{0}\right)\sinh4\sigma^{2}X+\left(1-V_{0}\right)\cosh4\sigma^{2}X\right].   
\end{equation}
Note, however that the curvature is no longer constant
\begin{equation}\label{d08}
\hat{R}=-\partial_{X}^{2}\hat{\xi}=-\left(1+V_{0}\right)\cosh 4\sigma^{2}X-\left(1-V_{0}\right)\sinh 4\sigma^{2}X.   
\end{equation}
In the limit $\sigma\rightarrow 0$, we obtain an $\mathrm{AdS}$ space for $V_0=1$ and a flat space for $V_{0}=-1$, in contrast with the potentials \eqref{d06} where we only obtain $\mathrm{AdS}$ space independently of the value of the parameter $\sigma$. Therefore, the hyperbolic potentials \eqref{d07} are conformally equivalent to the $\mathrm{AdS}_{2}$ models \eqref{d06} but they represent different physical realities.

\section{First-order formulation of dilaton gravity}\label{sec3}

For $\mathrm{AdS_2}$ spaces, the boundary conditions are proposed after solving the equations of motion in the Fefferman--Graham gauge \cite{Fefferman:2007rka}. This can be a difficult task especially if we consider fluctuating dilaton solutions. One way to avoid the second--order equations of motion from \label{d01} is to consider the first--order formulation of dilaton gravity. The first--order dilaton action is:
\begin{equation}\label{ap01}
I_{1st} = \frac{k}{2\pi}\int_{\mathcal{M}}\left[X^{a}\left(\mathrm{d}e_{a}+\epsilon_{a}^{\;b}\omega\wedge e_{b}\right)+X\mathrm{d}\omega-\frac{1}{2}\mathcal{V}\epsilon^{ab}e_{a}\wedge e_{b}\right]    
\end{equation}
where the Latin indices $a,b$ take values: $1,2$ and are raised and lowered with the Euclidean metric $\delta_{ab}=\mathrm{diag}\left(1,1\right)$. The Levi--Civita symbol is denoted by $\epsilon_{ab}$ ($\epsilon_{12}=1$ by convention). The Cartan variables are the zwebein $e_a$ and the (dualized) spin-connection $\omega$. In \eqref{ap01} we also introduced auxiliary fields $X^{a}$ which impose constraints on the torsion. The potential $\mathcal V$ is given by: 
\begin{equation}\label{ap02}
\mathcal{V}\left(X,X^{a}X_{a}\right)\equiv V+\frac{1}{2}X^{a}X^{b}\delta_{ab}U.    
\end{equation}
In \cite{Bergamin:2004pn} was shown that the first--order action \eqref{ap01} is equivalent to the second--order one \eqref{D01}  up to a boundary term
\begin{equation}\label{ap03}
I_{1st}	= I + \frac{k}{2\pi}\int\mathrm{d}^2x\;\partial_{\mu}\left(X\varepsilon^{\mu\nu}t_{\nu}\right)    
\end{equation}
where $t=Ue_{a}X^{a}$ is the contorsion. 

For models with $U\neq0$, i.e., models with non-vanishing contorsion, it is convenient to write the equations of motion in term of the (dualized) Levi-Civita connection $\Omega=\omega-t$:
\begin{eqnarray}
\mathrm{d}X &=& \epsilon_{\;b}^{a}X^{b}e_{a}\label{ap04a}\\
\mathrm{d}X^{a}	&=& -\epsilon^{ab}\left[\Omega X_{b}-Ve_{b}-\frac{1}{2}U\left(X^{c}X_{c}e_{b}-2X_{b}X^{c}e_{c}\right)\right]\label{ap04b}\\
\mathrm{d}e_{a}	&=& -\epsilon_{ab}\Omega\wedge e^{b}\label{ap04c}\\
\mathrm{d}\Omega &=& \frac{1}{2}e^{-Q}\left[\partial_{X}^{2}w-\left(\mathcal{C}-w\right)\partial_{X}U+U\partial_{X}w\right]\epsilon^{ab}e_{a}\wedge e_{b}.\label{ap04d}
\end{eqnarray}
From the equations of motion we can identify the Casimir: 
\begin{equation}
\mathcal{C}\equiv w+\frac{1}{2}e^{Q}X^{a}X^{b}\delta_{ab}    \label{ap05}
\end{equation}
a constant of motion $\mathrm{d}\mathcal C=0$ which was previously presented in \eqref{d04}.

For dilaton gravity in Euclidean signature it is also convenient to define complex fields \cite{Bergamin:2004pn}
\begin{equation}\label{c01}
Y\equiv\frac{1}{\sqrt{2}}\left(Y^{1}+iY^{2}\right),	\quad	\bar{Y}\equiv\frac{1}{\sqrt{2}}\left(Y^{1}-iY^{2}\right),\quad
e\equiv\frac{1}{\sqrt{2}}\left(e_{1}+ie_{2}\right),	\quad	\bar{e}\equiv\frac{1}{\sqrt{2}}\left(e_{1}-ie_{2}\right).
\end{equation}
In this formulation the complex fields are considered to be independent. Then, we can rewrite the action \eqref{ap01} in term of these new variables and compute the equations of motion. For the Cartan variables, these equations are:
\begin{eqnarray}
\mathrm{d}e	&=& i\Omega\wedge e  \label{c02a}\\
\mathrm{d}\bar{e}	&=& -i\Omega\wedge \bar {e}  \label{c02b}\\
\mathrm{d}\Omega &=& -ie^{-Q}\left[\partial_{X}^{2}w-\left(\mathcal{C}-w\right)\partial_{X}U+U\partial_{X}w\right]\bar{e}\wedge e. \label{c02c}
\end{eqnarray}
For the dilaton and auxiliary fields, the equations of motion are
\begin{eqnarray}
\mathrm{d}X	&=& i\bar{Y}e-iY\bar{e},\\
\mathrm{d}Y	&=& i\Omega Y-iVe+iUYY\bar{e},\\
\mathrm{d}\bar{Y} &=& -i\Omega \bar{Y}+iV\bar{e}-iU\bar{Y}\bar{Y}e.
\end{eqnarray}
From the dilaton equations we can also identify the Casimir: $\mathcal{C}\equiv w+e^{Q}Y\bar{Y}$.

We have explicitly written the equations for $Y,\,\bar Y$ and $e,\,\bar e$ to reinforce that these variables are independent. The only real variables are the Levi--Civita connection and the dilaton field. The first--order formulation of dilaton gravity, in real or complex variables, can be recast as a Poisson-sigma model \cite{Schaller:1994es}.  

\section{Boundary conditions in the Fefferman--Graham gauge} \label{sec4}

In this section we solve the first-order equations of motion \eqref{ap04a}-\eqref{ap04d} in order to propose boundary conditions for the $\mathrm{AdS_2}$ gravity models. Let us consider that the two-manifold $\mathcal M$ has the topology of a disk with coordinates $(\rho,\tau)$. The radial coordinate is $\rho$ and the boundary $\partial \mathcal M=S^1$ is located at $\rho=\rho_b \rightarrow \infty$. The boundary coordinate is $\tau$, with periodicity $\beta$: $\tau \sim \tau+\beta$. The line element in the Fefferman--Graham gauge is
\begin{equation}
    \mathrm{d}s^2 = \mathrm{d}\rho^2 + h^2\mathrm{d}\tau^2 \label{ap06a}
\end{equation}
where $h=h(\rho,\tau)$. Equivalently, we can choose the following ansatz for the zweibein 
\begin{equation}
e_{1\rho}=0,	\quad	e_{1\tau}=h,\quad e_{2\rho}=1,	\quad	e_{2\tau}=0,  \label{ap06}
\end{equation}
and from \eqref{ap04c} we obtain the components of the Levi-Civita spin-connection
\begin{equation}
\Omega_{\rho}=0,	\quad	\Omega_{\tau}=\partial_{\rho}h. \label{ap07}
\end{equation}
Replacing \eqref{ap06}, \eqref{ap07} in \eqref{ap04d} we obtain the simple equation $\partial^2_\rho h = h$. Let us consider the particular solution
\begin{equation}
h=e^\rho - \mathcal L(\tau) e^{-\rho}   
\end{equation}
with fixed leading order. This choice is common in holography, however, it can be generalized, as showed in \cite{Grumiller:2017qao}. With this solution we determine all components for the zweibein and Levi-civita spin-connection. We request that the Cartan variables approach to 
\begin{equation}
 e_{2\rho}=1,\qquad  e_{1\tau} = e^\rho - \mathcal L e^{-\rho},\qquad  \Omega_{\tau} =  e^\rho + \mathcal L e^{-\rho} \label{bc01}
\end{equation}
in the asymptotic region, i.e., at $\rho\rightarrow \infty$. These are our boundary conditions which, in the metric variables are equivalent to
\begin{equation}
    g_{\rho\rho}=1,\qquad g_{\rho\tau} = 0, \qquad g_{\tau\tau} = (e^\rho - \mathcal L e^{-\rho})^2. \label{bc02}
\end{equation}
It is quite remarkable that these boundary conditions are the same that the ones proposed for $\mathrm{JT}$ gravity in \cite{Grumiller:2015vaa}. 

Now, let us solve the radial components of the equations of motion for the dilaton and the auxiliary fields:
\begin{eqnarray}
\partial_{\rho}X	&=&-X_{1} \label{ap08a}\\
\partial_{\rho}X_{1} &=& \frac{1}{8\sigma^{2}}\left(V_{0}e^{-8\sigma^{2}X}-1\right)+2\sigma^{2}\left(X_{1}\right)^{2}-2\sigma^{2}\left(X_{2}\right)^{2} \label{ap08b}\\
\partial_{\rho}X_{2} &=& 4\sigma^{2}X_{1}X_{2}. \label{ap08c}
\end{eqnarray} 
Furthermore, the Casimir \eqref{ap05} is
\begin{equation}
\mathcal{C}=-\frac{c_{0}}{16\sigma^{4}}=-\frac{1}{32\sigma^{4}}\left(V_{0}e^{-4\sigma^{2}X}+e^{4\sigma^{2}X}\right)+\frac{1}{2}e^{4\sigma^{2}X}\left[\left(X_{1}\right)^{2}+\left(X_{2}\right)^{2}\right].    \label{ap09} 
\end{equation}
The strategy to solve the system of equations is the following: We use \eqref{ap09} to write $(X_{2})^2$ as function of the Casimir, the dilaton and $X_1$. We then introduce this expression in \eqref{ap08b}. Since $X_{1}$ is essentially the derivative of the dilaton we can obtain a second-order differential equation for the dilaton
\begin{equation}\label{ap10}
\partial_{\rho}^{2}X = \frac{1}{4\sigma^{2}}-4\sigma^{2}\left(\partial_{\rho}X\right)^{2}-\frac{c_{0}}{4\sigma^{2}}e^{-4\sigma^{2}X}. 
\end{equation}
Note the presence of the Casimir $c_{0}$ in this equations. It is convenient to define a transformed dilaton field $\tilde{X}$:
\begin{equation}
X \equiv \frac{1}{4\sigma^{2}}\ln\tilde{X}.    
\end{equation}
Then, dilaton equation \eqref{ap10} takes a very simple form
\begin{equation}
\partial_{\rho}^{2}\tilde{X}=\tilde{X}-c_{0}  \label{ap11}
\end{equation}
with solution 
\begin{equation}
\tilde{X}=x_{+}e^{\rho}+c_{0}+x_{-}e^{-\rho}   \label{ap12}  
\end{equation}
where $x_\pm$ and $c_0$ are functions of $\tau$.  We must prove, latter, that $c_{0}=0$ is conserved on-shell. 

From \eqref{ap08a} and \eqref{ap08c} we obtain the auxiliary fields $X^a$ in terms of the transformed dilaton $\tilde X$:
\begin{equation}
X_{1}=-\frac{1}{4\sigma^{2}}\tilde{X}^{-1}\partial_{\rho}\tilde{X},\qquad X_{2}=\tilde{X}^{-1}g \label{ap13}
\end{equation}
where $g=g\left(\tau\right)$ is an additional boundary function. From equation \eqref{ap12} we obtain the following asymptotic expansion for the auxiliary fields:
\begin{eqnarray}
X_{1} &=& -\frac{1}{4\sigma^{2}}+\frac{1}{4\sigma^{2}}c_{0}x_{+}^{-1}e^{-\rho}+\frac{1}{2\sigma^{2}}x_{-}x_{+}^{-1}e^{-2\rho}+\mathcal{O}\left(e^{-3\rho}\right)  \label{ap14a}\\
X_{2} &=& g \left[ x_{+}^{-1}e^{-\rho} - c_{0}x_{+}^{-2}e^{-2\rho} + \mathcal{O}\left(e^{-3\rho}\right)\right].\label{ap14b}
\end{eqnarray}
Equations \eqref{ap12}, \eqref{ap14a} and \eqref{ap14b} represent our boundary conditions for the dilaton and the auxiliary fields, respectively. 

The boundary equations of motion are obtained by replacing our boundary conditions in the $\tau$-components of equations \eqref{ap04a} and \eqref{ap04b}. We obtain
\begin{eqnarray}
g &=& \frac{1}{4\sigma^{2}}x'_{+},  \label{ap15a}\\
x'_{-} &=& -4\sigma^{2}g\mathcal{L},  \label{ap15b}\\ 
2\sigma^{2}g' &=& x_{-}-x_{+}\mathcal{L},  \label{ap15c}\\
c'_{0} &=& 0   \label{ap15d}
\end{eqnarray}
where the prime denotes derivative with respect to $\tau$. From equations \eqref{ap15a}-\eqref{ap15c} we obtain a Schwarzian derivative
\begin{equation}
    \frac{1}{2}x'''_+ + 2x'_+ \mathcal L + x_+ \mathcal L'=0 \label{ap16}
\end{equation}
which is the first hint of a $\mathrm{CFT}$. Equation \eqref{ap15d} states that the Casimir is conserved. Furthermore, by replacing the boundary conditions in the Casimir \eqref{ap09}, we obtain
\begin{equation}
c_{0}^{2}=4x_{+}x_{-}-16\sigma^{4}g^{2}+V_{0} = 2x_+ x''_+ + 4x^2_+ \mathcal L - (x'_+)^2 + V_0. \label{ap17}
\end{equation}
Note that the first three terms at the right hand side of the equation resemble the Casimir from $\mathrm{JT}$ gravity \cite{Grumiller:2017qao}. We can also obtain  \eqref{ap16} by taking the derivative of \eqref{ap17} and use the conservation of the Casimir.

Boundary conditions partially break the original symmetries of dilaton gravity, i.e., invariance under diffeomorphism (in the second--order 
formalism \eqref{D01}) or the non--linear gauge symmetry (in the first--order formalism \eqref{ap01}). Only a reduced group remains as symmetries, they 
are called asymptotic symmetries. Therefore, to obtain the asymptotic symmetries of the $\mathrm{AdS_2}$ gravity models, we need to know how the 
boundary conditions affect the original invariance under diffeomorphism:
\begin{eqnarray}
\delta_{\xi}g_{\mu\nu} &=& \xi^{\alpha}\partial_{\alpha}g_{\mu\nu}+g_{\mu\alpha}\partial_{\nu}\xi^{\alpha}+g_{\nu\alpha}\partial_{\mu}\xi^{\alpha} \label{ap18a}\\
\delta_{\xi}X &=& \xi^{\alpha}\partial_{\alpha}X \label{ap18b}
\end{eqnarray}
where $\xi^\mu$ are the diffeomorphism parameters. 

From the boundary condition \eqref{bc02} we obtain that the diffeomorphism parameters are
\begin{equation}
\xi^{\rho}=-\eta',\quad\xi^{\tau}=\eta-\frac{1}{2}\eta''e^{-2\rho}+\mathcal O\left(e^{-3\rho}\right). \label{ap19}
\end{equation}
By decomposing into Fourier modes, we obtain that these vectors close the Witt algebra. Furthermore, the field $\mathcal{L}$ transforms with an infinitesimal Schwarzian derivative:
\begin{equation}
\delta_{\eta}\mathcal{L}	=	\eta\mathcal{L}'+2\mathcal{L}\eta'+\frac{1}{2}\eta'''.    \label{ap20}
\end{equation}
where $\eta=\eta\left(\tau\right)$. Therefore, $\mathcal{L}$ behaves like a CFT stress tensor. 

Since $\tilde X$ also transforms as \eqref{ap18b}, we obtain the following transformations:
\begin{eqnarray}
\delta_{\eta}x_{+} &=& \eta x'_{+}-\eta'x_{+} \label{ap21a}\\
\delta_{\eta}c_{0} &=& \eta c'_{0} \label{ap21b}\\
\delta_{\eta}x_{-} &=& \eta'x_{-}+\eta x'_{-}-\frac{1}{2}\eta''x'_{+}. \label{ap21c}
\end{eqnarray}
The transformation for the dilaton leading order $x_+$ is similar to a boundary vector. The transformations for the Casimir $c_0$ and $x_-$ are similar to the JT case. Note that on-shell, the transformation for the Casimir is zero, since $c'_0=0$.  

\section{Boundary conditions in the Eddington--Finkelstein gauge} \label{sec5}

As we mentioned in the introduction, $\mathrm{AdS}_2$ holography can also be explored in the Eddington--Finkelstein gauge \cite{Godet:2020xpk,Afshar:2021qvi,Ruzziconi:2020wrb,Ecker:2021guy}. In this section we explore this gauge using the first--order formalism in complex variables. Let us begin writing the line element
\begin{equation}\label{EF01}
\mathrm{d}s^2 = 2i\mathrm{d}u\mathrm{d}r+2B\left(r,u\right)\mathrm{d}u^{2}
\end{equation}
where $r$ is the radial coordinate, $u$ is the retarded time and $B$ an arbitrary function. Equivalently, we can choose the following ansatz for the complex zweibein and Levi-Civita connection
\begin{equation}
e_{r}=0, \quad \bar{e}_{r}=i, \quad e_{u}=1, \quad \bar{e}_{u}=B, \quad \Omega_{r}=0,\quad \Omega_{u}=-\partial_{r}B. \label{EF02}
\end{equation}
The above choice is consistent with equations \eqref{c02a} and \eqref{c02b}. Replacing \eqref{EF02} in \eqref{c02c} for the $\mathrm{AdS}_2$ dilaton gravity models, we obtain the following equation: $\partial_{r}^{2}B=1$, with solution
\begin{equation}\label{EF03}
B\left( r,u \right) = \frac{1}{2}r^2 + \mathcal P\left( u \right)r + \mathcal T\left( u \right)
\end{equation}
where $\mathcal P$ and $\mathcal T$ are function of the retarded time. Since we have obtained the function $B$, we have now determined all components of the complex Cartan variables. We request that in the asymptotic region $r\rightarrow\infty$, the Cartan variables tend to \eqref{EF02} with $B$ given by \eqref{EF03}. These are our boundary condition in the Eddington--Finkelstein gauge.

Now we have to solve the dilaton equations:
\begin{eqnarray}
\mathrm{d}X	&=& i\bar{Y}e-iY\bar{e}  \label{EF04a}\\
\mathrm{d}Y	&=& i\Omega Y-\frac{1}{8\sigma^{2}}i\left(V_{0}e^{-8\sigma^{2}X}-1\right)e+4i\sigma^{2}YY\bar{e} \label{EF04b}\\
\mathrm{d}\bar{Y} &=& -i\Omega\bar{Y}+\frac{1}{8\sigma^{2}}i\left(V_{0}e^{-8\sigma^{2}X}-1\right)\bar{e}-4i\sigma^{2}\bar{Y}\bar{Y}e. \label{EF04c}
\end{eqnarray}
In Appendix \ref{AP} we solve the system for the particular case $V_0=1$ and $\sigma\rightarrow 0$, i.e, JT gravity. In general, for $\sigma\neq 0$ the radial components of \eqref{EF04a}-\eqref{EF04c} are
\begin{eqnarray}
\partial_{r}X	&=&	Y  \label{EF05a}\\
\partial_{r}Y	&=&	-4\sigma^{2}Y^2  \label{EF05b}\\
\partial_{r}\bar{Y}	&=&	-\frac{1}{8\sigma^{2}}\left(V_{0}e^{-8\sigma^{2}X}-1\right). \label{EF05c}
\end{eqnarray}
We can combine \eqref{EF05a} with \eqref{EF05b} to obtain
\begin{equation} 
\partial_{r}^{2}X+4\sigma^{2}\left(\partial_{r}X\right)^{2}	=	0. \label{EF06}     
\end{equation}
Note that this equation does not depend on the Casimir, in contrast with the dilaton equation in the Fefferman--Graham case \eqref{ap10}. In order to solve \eqref{EF06} we define $X=\frac{1}{4\sigma^{2}}\ln\Phi$. Then, the equation reduces to $\partial_{r}^{2}\Phi=0$, with solution
\begin{equation}
\Phi\left(r,u\right)=\varphi_{1}\left(u\right)r+\varphi_{0}\left(u\right) \label{EF07}
\end{equation}
where $\varphi_{1,0}$ are arbitrary functions of the retarded time. We can obtain $Y$ from equation \eqref{EF05a}:
\begin{equation}
Y = \frac{1}{4\sigma^{2}}\Phi^{-1}\partial_{r}\Phi = \frac{1}{4\sigma^{2}}\left[r^{-1}-\varphi_{0}\varphi_{1}^{-1}r^{-2}+\mathcal{O}\left(r^{-3}\right)\right]  \label{EF08}
\end{equation}
and, from \eqref{EF05c} we obtain the following equation for $\bar Y$:
\begin{equation}
\partial_{r}\bar{Y} = \frac{1}{8\sigma^{2}}-\frac{1}{8\sigma^{2}}V_{0}\Phi^{-2}=\frac{1}{8\sigma^{2}}-\frac{1}{8\sigma^{2}}V_{0}\varphi_{1}^{-2}r^{-2}\left(1-2\varphi_{1}^{-1}\varphi_{0}r^{-1}+...\right).     \label{EF09}
\end{equation}
This expression can be integrated
\begin{equation}
\bar{Y} = \frac{1}{8\sigma^{2}}\left[r + \ell\left(u\right) + V_{0}\varphi_{1}^{-2}r^{-1}+\mathcal{O}\left(r^{-2}\right)\right].     \label{EF10}
\end{equation}
where $\ell$ is a function of the retarded time, this function is introduced at the moment of integration. The expressions \eqref{EF08}-\eqref{EF10} represent our boundary conditions for the auxiliary complex fields. 

Since we have the boundary condition for the dilaton part, we can now compute the Casimir. In this case we obtain the following finite expression
\begin{equation}
c_{0} = -16\sigma^{4}\mathcal{C}=\frac{1}{2}\left(\varphi_{0}-\ell\varphi_{1}\right).  \label{EF12}
\end{equation}

From the  temporal components of \eqref{EF04a}-\eqref{EF04c} we obtain the boundary equations of motion
\begin{eqnarray}
i\dot{\varphi}_{1} &=& -\frac{1}{2}\varphi_{0}-\frac{1}{2}\varphi_{1}\ell+\mathcal{P}\varphi_{1} \label{EF13a}\\
i\dot{\varphi}_{0} &=& -\frac{1}{2}\varphi_{0}\ell+\mathcal{T}\varphi_{1}-\frac{1}{2}V_{0}\varphi_{1}^{-1} \label{EF13b}\\
i\dot{\ell} &=& \mathcal{T}+\frac{1}{2}\ell^{2}-\ell\mathcal{P}-\frac{1}{2}V_{0}\varphi_{1}^{-2} \label{EF13c}
\end{eqnarray}
where the dot denotes partial derivation with respect to the retarded time: $\partial_u$. From these equations we can show that the Casimir \eqref{EF12} is conserved $\partial_u c_0=0$. Furthermore, by a simple manipulation of the equations we can write the Casimir as a function of $\varphi_1$, $\mathcal P$, $\mathcal T$ and their derivatives:
\begin{equation}
c_{0}^{2}=2\varphi_{1}\ddot{\varphi}_{1}-\dot{\varphi}_{1}^{2}+4\varphi_{1}^{2}\left(\frac{1}{2}i\dot{\mathcal{P}}+\frac{1}{4}\mathcal{P}^{2}-\frac{1}{2}\mathcal{T}\right)+V_{0}.    \label{EF14}
\end{equation}
This expression is similar to \eqref{ap17} if we identify $\phi_1\rightarrow x_+$ and $\mathcal L \rightarrow \frac{1}{2}i\dot{\mathcal{P}}+\frac{1}{4}\mathcal{P}^{2}-\frac{1}{2}\mathcal{T}$. Therefore, if we derive \eqref{EF14} and use the conservation of the Casimir, we obtain a Schwarzian derivative for the field $\varphi_1$. For JT gravity, this identification can be derived by relating the solution space of Eddington--Finkelstein and Fefferman--Graham gauge.

In the Eddington--Finkelstein gauge the diffeomorphism parameters are:
\begin{equation}
\xi^{u} = \alpha, \qquad \xi^{r} = -\dot{\alpha}r + \beta    \label{EF15}
\end{equation}
where $\alpha$ and $\beta$ are independent functions of the retarded time. Performing a decomposition into Fourier modes, we notice that these vectors close a warped Witt algebra. Furthermore, they produce the following transformation on the fields $\mathcal P$, $\mathcal T$ and in the dilaton components
\begin{eqnarray}
\delta_{\xi}\mathcal{P} &=& \alpha\dot{\mathcal{P}}+\mathcal{P}\dot{\alpha}-i\ddot{\alpha}+\beta \label{EF16a}\\
\delta_{\xi}\mathcal{T} &=& \alpha\dot{\mathcal{T}}+2\mathcal{T}\dot{\alpha}+i \dot{\beta} + \mathcal{P}\beta \label{EF16b}\\
\delta_{\xi} \varphi_{1} &=&	\alpha\dot{\varphi}_{1}-\varphi_{1}\dot{\alpha}\\
\delta_{\xi} \varphi_{0} &=&	\varphi_{1}\beta+\dot{\varphi}_{0}\alpha.
\end{eqnarray}
By performing a shift $\mathcal P\rightarrow i\mathcal P$, $\beta \rightarrow i\beta$, $\varphi_0\rightarrow i\varphi_0$, we notice that the transformation of the fields $\mathcal P$ and $\mathcal T$ coincide with the ones of a warped Virasoro, whenever $\beta$ is a total derivative, and that the leading order of the dilaton $\Phi$ transforms as a boundary vector.  

\section{Thermodynamics} \label{sec6}

A good test of consistency for our boundary conditions is the computation of the thermodynamic properties of the black hole solutions. These solutions are obtained for static configurations after performing a Wick rotation. 

Let us begin with our solution in the Fefferman--Graham gauge where the static configuration is given by: $x_{+}=\bar{x}=cte$ and $\mathcal{L}=\bar{\mathcal{L}}=cte$. Performing a Wick rotation $\tau \rightarrow i\tau$ we notice that \eqref{ap06a} represents a black hole with a single horizon $\rho_h$ located at $e^{\rho_{h}} = \sqrt{ \bar{\mathcal L}}$. This solution is different from the one obtained in section \ref{sec2} which posses inner and outer horizon. 
Wald's formula states that the entropy is proportional to the value of the dilaton at the horizon. Then:
\begin{equation}
S_{\mathrm{Wald}}=k X\left(\rho_{h}\right)=\frac{k}{4\sigma^{2}}\ln\tilde{X}_h    \label{TH01}
\end{equation}
where $\tilde X_h$ denotes the value of the transformed dilaton at the horizon. For static configuration the boundary equations of motion reduce to 
\begin{equation}
x_{-}=\bar{\mathcal{L}}\bar{x},\qquad c_{0}^{2}=4\bar{\mathcal{L}}\bar{x}^{2}+V_{0}     \label{TH02}
\end{equation}
and, as a consequence, $\tilde X_h=c_{0}+2\bar{x}\sqrt{\bar{\mathcal{L}}}=c_{0}+\sqrt{c_{0}^{2}-V_{0}}$.  Therefore, the entropy \eqref{TH01} is given by
\begin{equation}
S_{\mathrm{Wald}}=\frac{k}{4\sigma^{2}}\ln\left(c_{0}+\sqrt{c_{0}^{2}-V_{0}}\right).    \label{TH03}
\end{equation}
This result coincides with our previous computation \eqref{entropy}.

We now follow a similar procedure for the solution in the Eddington--Finkelstein gauge. We perform a Wick rotation $u\rightarrow iu$ and consider static configuration: $\varphi_1=\bar \varphi=cte$, $\mathcal P=\bar{\mathcal P} = cte$ and $\mathcal T=\bar{\mathcal T} = cte$. This solution represents a black hole with outer horizon located at $r_h = -\bar{\mathcal P} + \left(\bar{\mathcal P}^2-2\bar{\mathcal T}\right)^{1/2}$. For this configuration, the boundary equations of motion are
\begin{equation}
\varphi_{0}	= c_{0} +\bar{\mathcal{P}}\bar{\varphi},\qquad
c_{0}^{2} = \bar{\varphi}^{2}\left(\mathcal{P}^{2}-2\mathcal{T}\right)+V_{0}.\label{TH04}
\end{equation}
Then, the value of the dilaton at the horizon is
\begin{equation}
X_h = \frac{1}{4\sigma^2} \ln\left[\bar{\varphi}r_h + \bar{\varphi}_0\right] = \frac{1}{4\sigma^2} \ln\left[\bar{\varphi}\left(-\bar{\mathcal P} + \sqrt{\bar{\mathcal P}^2-2\bar{\mathcal T}}\right) +  c_{0} +\bar{\mathcal{P}}\bar{\varphi}\right]       \label{TH05}
\end{equation}
Using the Casimir expression in \eqref{TH04}, we notice that \eqref{TH05} leads to the same Wald's entropy \eqref{TH03}. These results show that our boundary conditions in Fefferman--Graham and Eddington--Finkelstein gauge are consistent since they reproduce the correct entropy.

\section{Final remarks} \label{sec7}

In this work we proposed new boundary conditions for the $\mathrm{AdS_2}$ gravity models \cite{Hotta:1998iq}. We showed that for a constant potential $U$ this model can be built from \eqref{d05} as the most general dilaton gravity with constant curvature. If we repeat this procedure for flat space, we obtain a particular Liouville gravity model \cite{Bergamin:2004pn}: $U=4\sigma$ and $V=V_0 e^{-8\sigma X}$.

Our boundary conditions are different from the ones proposed in \cite{Hotta:1998iq}. The main difference is the gauge choice: In section \ref{sec4} we worked in the Fefferman--Graham gauge, which is the preferred gauge to study asymptotically $\mathrm{AdS}$ spacetimes, and in section \ref{sec5} we worked in the Eddington--Finkelstein  which can be adapted to asymptotically flat or asymptotically $\mathrm{AdS}$ spacetimes (see \cite{Ruzziconi:2019pzd}), while in \cite{Hotta:1998iq} the model \eqref{d06} was studied in the conformal gauge. Since we are dealing with a bi--parametric family of models with negative and constant curvature, our gauge choices are more natural for the study of $\mathrm{AdS}$ holography. In the Fefferman--Graham gauge we have a field $\mathcal L$ that belongs to the coadjoint orbit of the Virasoro group \eqref{ap20} and in the Eddington--Filkenstein gauge, we have two quantities $(\mathcal P,\mathcal T)$ that belong to the coadjoint representation of the warped Virasoro group \eqref{EF16a}, \eqref{EF16b}. These results are known for JT gravity and now we are generalizing them for the family of models \eqref{d06}. Another difference with \cite{Hotta:1998iq} is that we propose boundary conditions for the Cartan variables, the dilaton and auxiliary fields after solving the first--order equations of motion, rather than the second--order ones: This procedure allows generalizations, following a similar approach to \cite{Grumiller:2017qao}, leading to different sets of boundary conditions and consequently, to new set of asymptotic symmetries.  We also applied a consistency check to our boundary conditions: Since we are working in Euclidean signature, we computed the correct entropy of our black hole solutions from Wald's formula.

As future perspectives of this work, it would be interesting to explore if the entropy can be computed through a Cardy or Cardy--like formula. To achieve this we first need to compute the boundary charges. The integrability of the charges, in the Bondi gauge, has been studied in \cite{Ruzziconi:2020wrb} and applied for models with $U=0$. Then, it would be illustrative to check the integrability for models with constant $U$, such as the $\mathrm{AdS_2}$ models. 

Another topic that need further study is to obtain a Schwarzian action as the holographic dual of the $\mathrm{AdS_2}$ gravity models, i.e., as a 
result of computing the complete action $\Gamma=I + I_{\mathrm{b}}$ for fluctuating dilaton solution, where $I$ is the bulk action \eqref{D01} 
and $I_{\mathrm{b}}$ is a collection of boundary terms (see \cite{Grumiller:2007ju,Ecker:2021guy,Ruzziconi:2020wrb}).  For JT gravity and the 
$\mathrm{AdS}$ sector of the $ab$--family, the bulk action is zero and a boundary term, respectively. Then, the complete action is a boundary term 
proportional to the Casimir, and after some manipulations, we obtain a Schwarzian action. On the other hand, the bulk action for $\mathrm{AdS_2}$ 
gravity models is neither zero nor a boundary term for fluctuating dilaton solutions. We consider that for the $\mathrm{AdS_2}$ gravity models, the most appropriated approach to obtain the Schwarzian is exploring the target space diffeomorphism in the first--order formulation, as performed in \cite{Valcarcel:2022zqm} for the Rindler gravity model.

\section{Acknowledgements}

We thank D. Vassilevich for reading the manuscript and discussions. We also thank F. Ecker and D. Grumiller for their collaboration on two-dimensional gravity.

\appendix

\section{Boundary Conditions for JT gravity using complex fields} \label{AP}

Let us propose boundary conditions for the JT gravity case, i.e., $V_0=1$ and $\sigma\rightarrow 0$. The dilaton equations are
\begin{eqnarray}
\mathrm{d}X	&=&i\bar{Y}e-iY\bar{e} \label{JT01a}\\
\mathrm{d}Y	&=&	i\Omega Y + iXe \label{JT01b}\\
\mathrm{d}\bar{Y} &=&-i\Omega\bar{Y}-iX\bar{e}. \label{JT01c}
\end{eqnarray}
Replacing the conditions \eqref{EF02} and \eqref{EF03} in the radial components of equations \eqref{JT01a}-\eqref{JT01c} we obtain
\begin{equation} 
\partial_{r}X=Y,\quad\partial_{r}Y=0,\quad\partial_{r}\bar{Y}=X.     \label{JT02}
\end{equation}
This system of equations has solution
\begin{equation}
X=\phi_{1}r+\phi_{0},\quad Y=\phi_1,\quad\bar{Y}=\frac{1}{2}\phi_{1}r^{2}+\phi_{0}r+\phi_{-1}    \label{JT03}
\end{equation}
where $\phi_{0,\pm 1}$ are functions of the retarded time $u$. The equations \eqref{JT03} represent the boundary conditions for the dilaton and auxiliary complex fields. Replacing these conditions in the temporal components of the dilaton equation \eqref{JT01a}-\eqref{JT01c}, we obtain the following boundary equations of motion
\begin{eqnarray}
i\dot{\phi}_{1}	&=&	\phi_{1}\mathcal{P}-\phi_{0}\\
i\dot{\phi}_{0}	&=&	\phi_{1}\mathcal{T}-\phi_{-1}\\
i\dot{\phi}_{-1}	&=&	\phi_{0}\mathcal{T}-\mathcal{P}\phi_{-1}.
\end{eqnarray}
For JT gravity, the Casimir is given by $\mathcal C = -\frac{1}{2}X^{2}+Y\bar{Y}$. Replacing our boundary conditions \eqref{JT03} we obtain that $\mathcal{C}=\phi_{1}\phi_{-1}-\frac{1}{2}\phi_{0}^{2}$, and it can be easily checked that the Casimir is conserved $\partial_u \mathcal C=0$. The fields $\phi_0$ and $\phi_{-1}$ can be written in function of $\phi_1$, $\mathcal P$, $\mathcal T$ and their derivatives. Then, the Casimir takes the following form
\begin{equation}
\mathcal{C}=\frac{1}{2}\dot{\phi}_{1}^{2}-\phi_{1}^{2}\left(\frac{1}{2}\mathcal{P}^{2}+i\dot{\mathcal{P}}-\mathcal{T}\right)-\phi_{1}\ddot{\phi}_{1}.    
\end{equation}
Deriving this expression we obtain a Schwarzian derivative for $\phi_1$. The boundary conditions are equivalent to the ones presented in \cite{Godet:2020xpk} (in the second-order formulation).

\bibliographystyle{fullsort.bst}

\bibliography{main} 

\end{document}